\documentclass[twocolumn,english]{revtex4-1}
\usepackage[T1]{fontenc}
\usepackage[latin9]{inputenc}
\usepackage{amsbsy}
\usepackage{amstext}
\usepackage{graphicx}

\makeatletter

\newcommand*\LyXThinSpace{\,\hspace{0pt}}

\usepackage{hyperref}

\makeatother

\usepackage{babel}
\begin{document}
\title{Scanning-probe and information-concealing machine learning intermediate
hexatic phase and critical scaling of solid-hexatic phase transition
in deformable particles}
\author{Wei-chen Guo}
\affiliation{Guangdong Provincial Key Laboratory of Quantum Engineering and Quantum
Materials, School of Physics and Telecommunication Engineering, South
China Normal University, Guangzhou 510006, China}
\affiliation{Guangdong-Hong Kong Joint Laboratory of Quantum Matter, South China
Normal University, Guangzhou 510006, China}
\author{Bao-quan Ai}
\email{aibq@scnu.edu.cn}

\affiliation{Guangdong Provincial Key Laboratory of Quantum Engineering and Quantum
Materials, School of Physics and Telecommunication Engineering, South
China Normal University, Guangzhou 510006, China}
\affiliation{Guangdong-Hong Kong Joint Laboratory of Quantum Matter, South China
Normal University, Guangzhou 510006, China}
\author{Liang He}
\email{liang.he@scnu.edu.cn}

\affiliation{Guangdong Provincial Key Laboratory of Quantum Engineering and Quantum
Materials, School of Physics and Telecommunication Engineering, South
China Normal University, Guangzhou 510006, China}
\affiliation{Guangdong-Hong Kong Joint Laboratory of Quantum Matter, South China
Normal University, Guangzhou 510006, China}
\begin{abstract}
We investigate the two-dimensional melting of deformable polymeric
particles with multi-body interactions described by the Voronoi model.
We report machine learning evidence for the existence of the intermediate
hexatic phase in this system, and extract the critical exponent $\nu\approx0.65$
for the divergence of the correlation length of the associated solid-hexatic
phase transition. Moreover, we clarify the discontinuous nature of
the hexatic-liquid phase transition in this system. These findings
are achieved by directly analyzing system's spatial configurations
with two generic machine learning approaches developed in this work,
dubbed ``scanning-probe'' via which the possible existence of intermediate
phases can be efficiently detected, and ``information-concealing''
via which the critical scaling of the correlation length in the vicinity
of generic continuous phase transition can be extracted. Our work
provides new physical insights into the fundamental nature of the
two-dimensional melting of deformable particles, and establishes a
new type of generic toolbox to investigate fundamental properties
of phase transitions in various complex systems.
\end{abstract}
\maketitle
\emph{Introduction}.---The nature of two-dimensional (2D) melting
\citep{Strandburg_RMP_1988_2Dmelting} is delicate and can show non-trivial
dependence on several properties of the specific systems, e.g., softness
\citep{Ciamarra_PRM_2018,Durand_PRL_2019}, activity \citep{Ciamarra_Soft_Matter_2020,Komatsu_PRX_2015,Digregorio_PRL_2018,Krauth_Nat_Commun_2018,Krauth_JChemPhys_2019},
density \citep{Ning_Xu_PRL_2016}, potential \citep{Krauth_PRL_2015,Yilong_Han_Nature_2016,Ciamarra_PRL_2020,Hajibabaei_PRE_2019,Frenkel_PRL_1995},
pinning of particles \citep{Lowen_PRL_2013_pinning}, shape of particles
\citep{Glotzer_PRX_2017,Cugliandolo_PRL_2017}, topological constraints
\citep{Guerra_Nature_2018}, etc. Three different types of 2D melting
scenarios, namely, the one-step melting scenario \citep{Nelson_PRB_1978},
the hard-disk-like scenario \citep{Krauth_PRL_2011,Russo_PRL_2017,Thorneywork_PRL_2017},
and the Kosterlitz-Thouless-Halperin-Nelson-Young scenario \citep{Kosterlitz_JPhysC_1973,Halperin_PRL_1978,Halperin_PRB_1979,Young_PRB_1979}
have been identified. However, some questions are still left open.
The most important ones include the existence of the intermediate
hexatic phase and the fundamental nature of its associated phase transitions.

To solve these open questions, one crucial step is to identify different
phases in the relevant systems. This is usually done via investigating
the spatial decay of correlation functions of the translational order
and the bond-orientational order, combined with the unbinding behavior
of dislocations and disclinations \citep{Kosterlitz_JPhysC_1973,Halperin_PRL_1978,Halperin_PRB_1979,Young_PRB_1979}.
More specifically, it is suggested that the solid-hexatic phase transition
is associated with the disappearance of the quasi-long range translational
order and the increasing number of dislocations \citep{Ciamarra_PRM_2018},
and that the hexatic-liquid phase transition is associated with the
disappearance of the long range bond-orientational order and the dissociation
of the dislocation into disclinations \citep{Ciamarra_PRM_2018}.
However, a firm confirmation within this approach is still difficult,
partially due to the possibly enormous value of the hexatic correlation
length \citep{Ciamarra_PRL_2020} and also the fact that other complicated
defects, such as vacancies and grain boundaries, might appear near
the phase transitions \citep{Digregorio_arXiv_2021,Weikai_Qi_Soft_Matter_2014,Frenkel_PRE_2000}.
Noticing that machine learning techniques have emerged in recent years
as an efficient tool to investigate various problems on phase transitions
\citep{Broecker_arXiv_2017,Melko_Nat_Phys_2017,van_Nieuwenburg_Nat_Phys_2017,Broecker_Sci_Rep_2017,Lei_Wang_PRB_2016,Lee_PRE_2019,Shinjo_PRB_2020,Kottmann_PRL_2020,Rzadkowski_NJP_2020,Guo_arXiv_2020,Guo_PRE_2021,van_Nieuwenburg_arXiv_2021,Huaping_Li_PNAS_2021,Koch_Janusz_Nat_Phys_2018,Koch_Janusz_PRE_2021,Melko_PRB_2019,Zhenyu_Li_PRB_2019,Wanzhou_Zhang_PRE_2019,Bachtis_PRE_2020},
this thus raises the intriguing opportunity to develop new tools based
on these powerful techniques to reveal new physical insights into
these open questions, in particular, the ones concerning the existence
of the intermediate hexatic phase and the fundamental nature of its
associated phase transitions, especially the possible critical scaling
behavior.

\begin{figure}
\noindent \begin{centering}
\includegraphics[width=3.2in]{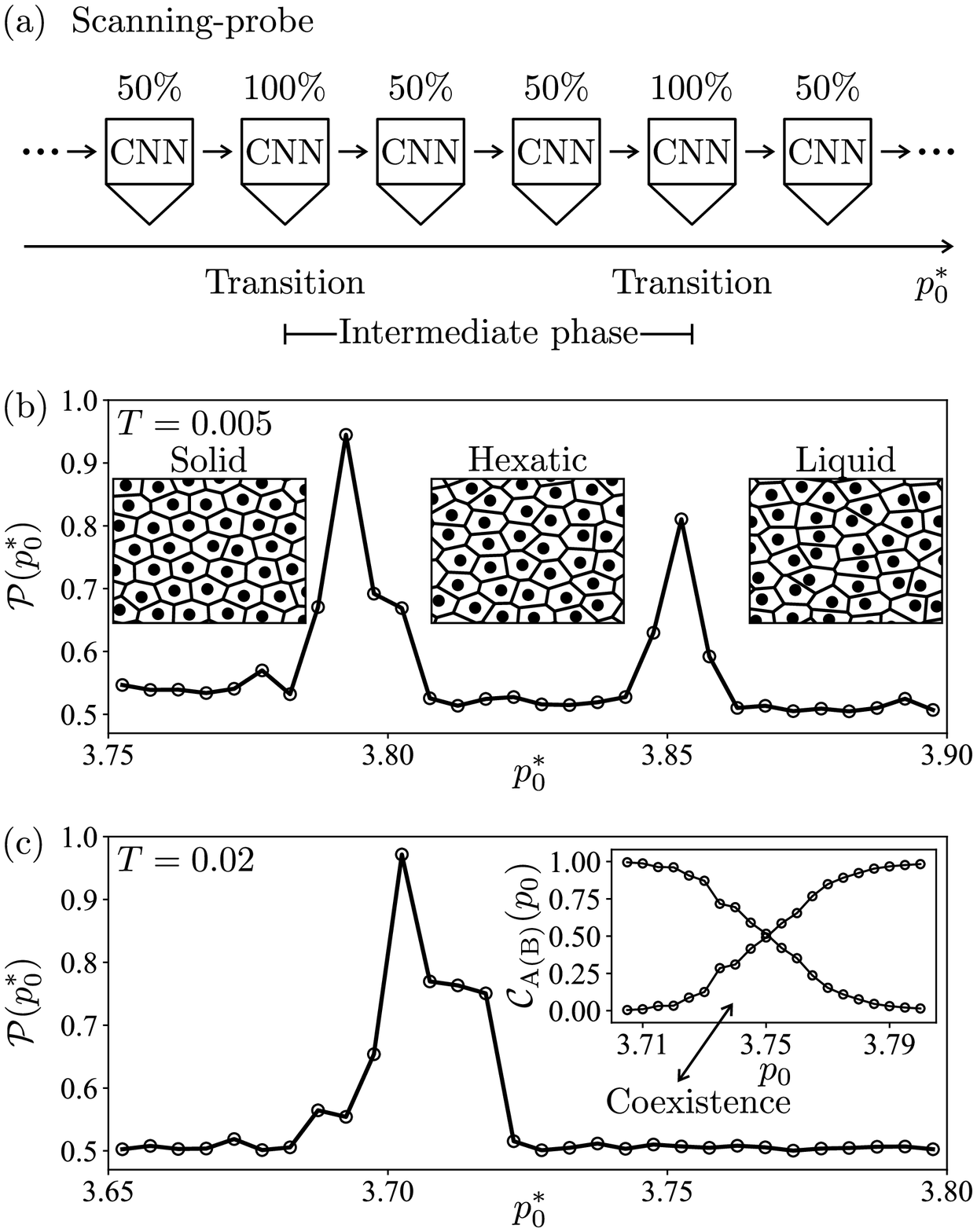}
\par\end{centering}
\caption{\label{fig:Identification-of-hexatic}(a) Schematic illustration of
``scanning-probe'' machine learning. The ``phase-transition-probe''
is realized by a deep CNN that is able to give the signal $\mathcal{P}(p_{0}^{*})$
on how likely there is a phase transition in a narrow parameter interval
$[p_{0}^{*}-\Delta p_{0},p_{0}^{*}+\Delta p_{0}]$, where $\mathcal{P}(p_{0}^{*})\approx100\%$
indicates the existence of a possible transition, with $p_{0}^{*}$
being the suspected phase transition point and $\Delta p_{0}$ being
the resolution of the probe. (b) Identification of the intermediate
hexatic phase at a low temperature $T=0.005$. The classification
accuracy $\mathcal{P}(p_{0}^{*})$ associated with the ``phase-transition-probe''
manifests two pronounced peaks that signify two phase transitions
and hence three distinct phases, with the intermediate one corresponding
to the hexatic phase. And these double peaks correspond to the solid-hexatic
phase transition point $p_{0}^{\textrm{SH}}=3.7925\pm0.0025$ and
the hexatic-liquid phase transition point $p_{0}^{\textrm{HL}}=3.8525\pm0.0025$,
respectively. Inset: Typical samples of the data in the solid phase
(left), the hexatic phase (middle), and the liquid phase (right).
(c) Identification of the intermediate hexatic phase at a relatively
high temperature $T=0.02$. The classification accuracy $\mathcal{P}(p_{0}^{*})$
associated with ``phase-transition-probe'' manifests only a single
peak in this case, and this peak corresponds to the solid-hexatic
phase transition point with $p_{0}^{\textrm{SH}}=3.7025\pm0.0025$.
Inset: The auxiliary machine learning to resolve the hexatic-liquid
phase coexistence. Two average confidence values $\mathcal{C}_{\textrm{A}(\textrm{B})}(p_{0})$
of the deep CNN show big differences around $100\%$ in the vicinities
of both ends of the parameter interval $[3.705,3.800]$ for $p_{0}$,
clearly indicating that there exist two distinct phases around $p_{0}=3.705$
and $p_{0}=3.800$, i.e., the hexatic phase and the liquid phase.
The parameter region around the intersection point is expected to
correspond to a phase coexistence region of these two phases, and
the upper bound for the hexatic phase region can be estimated by this
point with $p_{0}^{\textrm{H,UB}}=3.7507$. See text for more details.}
\end{figure}

\begin{figure}
\noindent \begin{centering}
\includegraphics[width=3.2in]{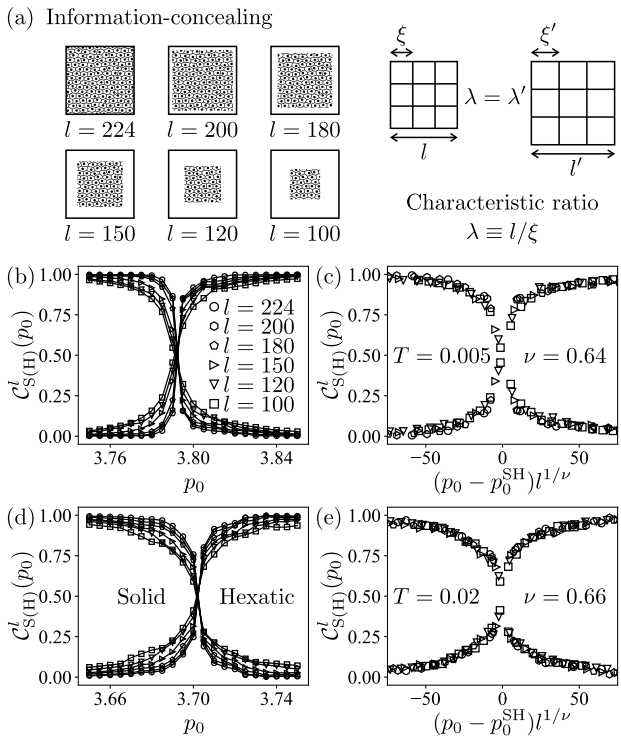}
\par\end{centering}
\caption{\label{fig:Critical-scaling}(a) Schematic illustration of ``information-concealing''
machine learning. The peripheries of the samples in the original dataset
are deliberately concealed to generate a series of datasets with reduced
effective areas characterized by the effective length scale $l$ for
their samples. The ratio between $l$ and the system correlation length
$\xi(p_{0})$, i.e, $\lambda=l\slash\xi(p_{0})$, determines the information
content of the system that can be utilized by the CNN to perform the
classification task. For instance, a small sample with linear size
$l$ and a correlation length $\xi(p_{0})$ is expected to contain
the same amount of information for the CNN as a larger sample with
linear size $l^{\prime}$ and a larger correlation length $\xi(p_{0}^{\prime})$
if $l\slash\xi(p_{0})=l^{\prime}\slash\xi(p_{0}^{\prime})$. This
indicates that the critical exponent $\nu$ for the divergence of
the correlation length can be extracted via the data collapse behavior
of the average confidence values that performs the classification
task with respect to datasets with samples of effective length scale
$l$. (b) Average confidence values of the CNN to classify the hexatic
phase from the solid phase using datasets with different fixed $l$
at the low temperature $T=0.005$. (c) The good data collapse is achieved
for the average confidence values in (b) as a function of $(p_{0}-p_{0}^{\textrm{SH}})l^{1\slash\nu}$
with the critical exponent $\nu=0.64$. (d) Average confidence values
of the CNN to classify the hexatic phase from the solid phase using
datasets with different fixed $l$ at the relatively high temperature
$T=0.02$. (e) The good data collapse is achieved for the average
confidence values in (d) as a function of $(p_{0}-p_{0}^{\textrm{SH}})l^{1\slash\nu}$
with the critical exponent $\nu=0.66$. See text for more details.}
\end{figure}

In this work, we address these questions for the 2D melting of deformable
polymeric particles with multi-body interactions described by the
Voronoi model \citep{Ciamarra_PRM_2018,Manning_PRL_2018,Manning_Nat_Phys_2015,Manning_PRX_2016,Marchetti_Soft_Matter_2018,Sussman_CPC_2017,Henkes_Nat_Commun_2020}.
To this end, we develop two generic neural network-based machine learning
approaches dubbed ``scanning-probe'' {[}cf.~Fig.~\ref{fig:Identification-of-hexatic}(a){]}
and ``information-concealing'' {[}cf.~Fig.~\ref{fig:Critical-scaling}(a){]}
to directly analyze a large number of system's spatial configurations
that are generated by Brownian dynamics simulations, and find the
following.

(i) Machine learning evidence for the existence of the intermediate
hexatic phase (cf.~Fig.~\ref{fig:Identification-of-hexatic}). At
the low temperature, when the so-called target shape index of the
system increases, the classification accuracy {[}cf.~Eq.~(\ref{eq:classification-accuracy}){]}
associated with the ``phase-transition-probe'' manifests two pronounced
peaks that signify two phase transitions and hence three distinct
phases, with the intermediate one corresponding to the hexatic phase
{[}cf.~Fig.~\ref{fig:Identification-of-hexatic}(b){]}. At the relatively
high temperature, the single peak behavior of the classification accuracy
of the ``phase-transition-probe'', combined with the phase coexistence
region identified by an auxiliary machine learning, still indicates
the existence of three distinct phases with the intermediate one corresponding
to the hexatic phase {[}cf.~Fig.~\ref{fig:Identification-of-hexatic}(c){]}.
The identified phase coexistence region also clarifies the discontinuous
nature of the hexatic-liquid phase transition in this system.

(ii) The continuous solid-hexatic phase transition manifests a critical
scaling behavior with the critical exponent $\nu\approx0.65$ for
the divergence of the correlation length (cf.~Fig.~\ref{fig:Critical-scaling}).
We observe that at different temperatures, the good data collapse
is achieved for the average outputs of the convolutional neural network
(CNN) employed in ``information-concealing'' machine learning as
a function of the reduced target shape index when it is rescaled with
the critical exponent $\nu\approx0.65$ {[}cf.~Figs.~\ref{fig:Critical-scaling}(c),~\ref{fig:Critical-scaling}(e){]}
with respect to the linear size $l$ of the system after ``concealing'',
clearly manifesting the critical scaling behavior of the correlation
length in the vicinity of the continuous solid-hexatic phase transition.

\emph{System and model}.---The system under study consists of $N$
deformable polymeric particles with multi-body interactions \citep{Ciamarra_PRM_2018},
whose elastic energy $E$ is modeled according to the Voronoi description
\citep{Manning_PRL_2018,Manning_Nat_Phys_2015,Manning_PRX_2016,Marchetti_Soft_Matter_2018,Sussman_CPC_2017,Henkes_Nat_Commun_2020},
\begin{equation}
E(\{\boldsymbol{r}_{i}\})=\sum_{i=1}^{N}[K_{A}(A_{i}-A_{0})^{2}+K_{P}(P_{i}-P_{0})^{2}],\label{eq:energy}
\end{equation}
where $A_{i}$ and $P_{i}$ are the cross-sectional area and the perimeter
of the $i$th particle, $A_{0}$ and $P_{0}$ are the preferred particle
area and perimeter, $K_{A}$ and $K_{P}$ represent the area and perimeter
stiffness moduli, respectively. This Voronoi model originates from
the investigations of the confluent cells in biological tissues, where
the first term in Eq.~(\ref{eq:energy}) models the cell incompressibility
and the monolayer's resistance to height fluctuations, and the second
term arises from the active contractility of the actomyosin subcellular
cortex and the effective cell membrane tension due to the cell-cell
adhesion and cortical tension. The effective dimensionless target
shape index $p_{0}=P_{0}/\sqrt{A_{0}}$ is an important parameter
that controls the elastic behavior of these deformable particles.
The elastic energy in Eq.~(\ref{eq:energy}) leads to a mechanical
interaction force on the $i$th particle by $\boldsymbol{F}_{i}=-\nabla_{i}E(\{\boldsymbol{r}_{i}\})$,
which is a multi-body interaction force that cannot be expressed as
a sum of pairwise force between the $i$th particle and its neighbors
\citep{Manning_PRL_2018,Manning_Nat_Phys_2015,Manning_PRX_2016,Marchetti_Soft_Matter_2018,Sussman_CPC_2017,Henkes_Nat_Commun_2020}.
To simulate the dynamics in this system, each particle undergoes overdamped
Brownian motion at a given temperature $T$. Thus, the dynamics of
the deformable particles in the overdamped limit follows the Langevin
equation \citep{Manning_PRL_2018,Manning_Nat_Phys_2015,Manning_PRX_2016,Marchetti_Soft_Matter_2018,Sussman_CPC_2017,Henkes_Nat_Commun_2020}

\begin{equation}
\frac{d\boldsymbol{r}_{i}}{dt}=\mu\boldsymbol{F}_{i}+\sqrt{2\mu k_{B}T}\boldsymbol{\xi}_{i}(t),\label{eq:Langevin}
\end{equation}
where $\mu$ is the mobility, $k_{B}$ is the Boltzmann constant,
and $\boldsymbol{\xi}_{i}(t)$ are Gaussian white noises with zero
mean and unit variance. Equations (\ref{eq:energy}),~(\ref{eq:Langevin})
can be rewritten in the dimensionless forms by introducing the characteristic
length $\sqrt{A_{0}}$ and time $1\slash(\mu K_{A}A_{0})$. The parameters
in the dimensionless forms can be rewritten as $\hat{K}_{P}=K_{P}/(K_{A}A_{0})$
and $\hat{T}=k_{B}T/(K_{A}A_{0}^{2})$. We shall omit the hat for
all quantities occurring in the above equations from now on and use
only the dimensionless variables.

From the expression for the elastic energy in Eq.~(\ref{eq:energy}),
we can see that at fixed temperature $T$, the system softens when
the target shape index $p_{0}$ increases. Therefore, one naturally
expects a solid phase with small $p_{0}$, and a liquid phase with
large $p_{0}$. With an intermediate $p_{0}$, one so-called hexatic
phase could possibly emerge as an intermediate phase between the solid
and the liquid phases. Recent investigations \citep{Ciamarra_PRM_2018,Durand_PRL_2019,Ciamarra_Soft_Matter_2020,Manning_PRL_2018,Manning_Nat_Phys_2015,Manning_PRX_2016,Marchetti_Soft_Matter_2018,Sussman_CPC_2017,Henkes_Nat_Commun_2020}
have shown numerical evidence that supports the existence of the intermediate
hexatic phase in this system from properties of the translational
correlation function, the bond-orientational correlation function,
and different types of topological defects such as dislocations and
disclinations. However, potential limitations of the existing evidence
were also reported, for instance, the exponent of the asymptotic power
law decay of the correlation functions can be difficult to extract
due to the limited system size employed in the study \citep{Komatsu_PRX_2015,Ciamarra_PRL_2020},
and the relevant types of topological defects that drive the phase
transition associated with the intermediate hexatic phase are still
under debate \citep{Digregorio_arXiv_2021,Weikai_Qi_Soft_Matter_2014,Frenkel_PRE_2000}.
In this regard, a type of approaches that directly analyzes the system's
spatial configurations with as few build-in empirical assumptions
as possible is highly desirable to provide new physical insights into
the questions concerning the existence of the intermediate hexatic
phase and the fundamental nature of its associated phase transitions
in this system. Indeed, as we shall see in the following sections,
such a type of approaches can be developed by utilizing the modern
machine learning techniques \citep{Nielsen_Book_2015,Goodfellow_Book_2016}.

Here in this work, we focus on the case with $K_{P}=1$, and set the
particle number $N=3960$ (cf.~Supplemental Material \citep{Supplemental_Material}
for an investigation on the finite size effect, which shows that it
does not impose strong influences on the major results presented in
the following). The configurations to be directly processed by the
machine learning approaches developed in the following are obtained
by Brownian dynamics simulations of the system in a 2D rectangular
space with aspect ratio $2\slash\sqrt{3}$ and periodic boundary condition
imposed. The initial configuration is chosen to be a perfect hexagonal
lattice. With different fixed values for $(T,p_{0})$, we typically
generate $2\times10^{3}$ of system's spatial configurations in the
steady state, and transform these configurations into images according
to the Voronoi tessellation \citep{Manning_PRL_2018,Manning_Nat_Phys_2015,Manning_PRX_2016,Marchetti_Soft_Matter_2018,Sussman_CPC_2017,Henkes_Nat_Commun_2020}
(see Supplemental Material \citep{Supplemental_Material} for more
technical details on the data generation).

\emph{Intermediate hexatic phase identified via ``scanning-probe''}.---To
identify the intermediate hexatic phase, we need a high-resolution
machine learning approach to efficiently detect multiple phase transitions
triggered by tuning a single system parameter, in contrast to most
of the well-established machine learning approaches that are experts
in searching for a single phase transition in a large parameter region
\citep{Melko_Nat_Phys_2017,van_Nieuwenburg_Nat_Phys_2017}. Here,
we develop a generic machine learning approach dubbed ``scanning-probe'',
in which the key workhorse is a ``phase-transition-probe'' that
is able to give the signal on how likely there is a phase transition
in a narrow parameter interval $[p_{0}^{*}-\Delta p_{0},p_{0}^{*}+\Delta p_{0}]$
with $p_{0}^{*}$ being the suspected phase transition point and $\Delta p_{0}$
being the resolution of the probe. Since the hexatic phase can be
far from obvious in the real space {[}cf.~insets of Fig.~\ref{fig:Identification-of-hexatic}(b){]},
the ``scanning-probe'' approach can be realized by a deep CNN that
has strong feature extraction ability, where the network is trained
as a binary classifier by using the data associated with the two boundaries
of the interval. More specifically, we label all the samples with
$p_{0}=p_{0}^{*}-\Delta p_{0}$ as ``phase A'' and the ones with
$p_{0}=p_{0}^{*}+\Delta p_{0}$ as ``phase B'', and the probe's
signal concerning the phase transition is given by the classification
accuracy $\mathcal{P}(p_{0}^{*})$ in the testing process of the trained
CNN, whose explicit form reads
\begin{equation}
\mathcal{P}(p_{0}^{*})\equiv\frac{M_{\textrm{A}}^{-}+M_{\textrm{B}}^{+}}{2M},\label{eq:classification-accuracy}
\end{equation}
with $M$ being the number of test samples associated with each $p_{0}$
and $M_{\textrm{A}}^{-}$ ($M_{\textrm{B}}^{+}$) being the number
of test samples that have been identified successfully as phase A
(B) by the CNN. Here, the CNN gives two outputs $(y_{\textrm{A}},y_{\textrm{B}})$,
where $y_{\textrm{A}(\textrm{B})}$ can be regarded as a classification
confidence value $\in[0,1]$ of how likely the sample is in phase
A (B). For a sample fed to the CNN, it is identified as phase A if
$y_{\textrm{A}}>y_{\textrm{B}}$ or phase B if $y_{\textrm{A}}<y_{\textrm{B}}$.
As one can naturally expect if the two groups of data belong to the
same phase physically, there should be no essential difference between
each other, hence the CNN can only learn to trivially identify all
the samples as the same phase. This leads to either $(M_{\textrm{A}}^{-},M_{\textrm{B}}^{+})\approx(M,0)$
or $(M_{\textrm{A}}^{-},M_{\textrm{B}}^{+})\approx(0,M)$, i.e., $\mathcal{P}(p_{0}^{*})\approx50\%$,
which is equivalent to a random guess. On the contrary, if there indeed
exists a phase transition in the interval $[p_{0}^{*}-\Delta p_{0},p_{0}^{*}+\Delta p_{0}]$,
the two groups of data can naturally manifest certain distinct differences
from each other that can be learned by the CNN, hence the classification
accuracy is expected to be very high, i.e., $\mathcal{P}(p_{0}^{*})\approx100\%$,
which can be used as a clear signal for the phase transition point
(see Supplemental Material \citep{Supplemental_Material} for more
technical details on the network training).

Via making a scanning with the ``phase-transition-probe'' over the
complete parameter region of interests, possible phase transitions
in it can be detected. Since each probing involves only two groups
of data that correspond to $p_{0}^{*}-\Delta p_{0}$ and $p_{0}^{*}+\Delta p_{0}$,
the computational cost of the ``scanning-probe'' approach scales
linearly with respect to the number of the suspected parameter points
in the complete parameter region. This is in contrast to the machine
learning approaches that employ all the data, e.g., the confusion
scheme \citep{van_Nieuwenburg_Nat_Phys_2017}, whose the computational
cost scales quadratically with respect to the number of the suspected
parameter points.

We first consider a nearly zero-temperature case at $T=0.005$. As
we can see from Fig.~\ref{fig:Identification-of-hexatic}(b), the
$p_{0}^{*}$ dependence of $\mathcal{P}(p_{0}^{*})$ manifests two
pronounced peaks, indicating that there should exist three distinct
phases in this system. From the expression of the system's elastic
energy in Eq.~(\ref{eq:energy}), we know that the system becomes
harder as the target shape index $p_{0}$ becomes smaller, hence the
system with a very small target shape index $p_{0}$ is expected to
be in the solid phase. Combining this physical understanding with
the behavior that $\mathcal{P}(p_{0}^{*})\approx50\%$ on the left
side of the left peak in Fig.~\ref{fig:Identification-of-hexatic}(b),
which manifests that the system's spatial configurations associated
with the small interval $[p_{0}^{*}-\Delta p_{0},p_{0}^{*}+\Delta p_{0}]$
show no significant difference among them, we can conclude that the
system is solid if $p_{0}$ is smaller than the target shape index
value corresponding to the left peak. In contrast, the system with
a very large $p_{0}$ should be in the liquid phase, since the increase
of $p_{0}$ favors less compact shapes and thus a reduction in the
number of sides of the Voronoi tessellation \citep{Ciamarra_PRM_2018,Durand_PRL_2019,Ciamarra_Soft_Matter_2020,Manning_PRL_2018,Manning_Nat_Phys_2015,Manning_PRX_2016,Marchetti_Soft_Matter_2018,Sussman_CPC_2017,Henkes_Nat_Commun_2020}.
Combining with the fact that $\mathcal{P}(p_{0}^{*})\approx50\%$
on the right side of the right peak in Fig.~\ref{fig:Identification-of-hexatic}(b),
we can conclude that the system is liquid if $p_{0}$ is larger than
the target shape index value corresponding to the right peak. In this
regard, one can naturally arrive at a further conclusion that an intermediate
phase exists between the two peaks, which is neither the solid phase
nor the liquid phase, and is expected to be the intermediate hexatic
phase that emerges between the solid and the liquid phases in 2D systems.
Consequently, the left peak in Fig.~\ref{fig:Identification-of-hexatic}(b)
corresponds to the solid-hexatic phase transition point with $p_{0}^{\textrm{SH}}=3.7925\pm0.0025$,
while the right one corresponds to the hexatic-liquid phase transition
point with $p_{0}^{\textrm{HL}}=3.8525\pm0.0025$, which is consistent
with the phase transition points $p_{0}^{\textrm{SH}}\approx3.78$
and $p_{0}^{\textrm{HL}}\approx3.85$ at $T=0.005$ identified via
conventional approaches \citep{Ciamarra_PRM_2018}, providing strong
machine learning evidence for the existence of the intermediate hexatic
phase.

We further investigate a relatively high temperature case at $T=0.02$.
In this case, the deep CNN detects only a single peak, as shown in
Fig.~\ref{fig:Identification-of-hexatic}(c). This seems to imply
that there exist only two phase at this temperature. However, we note
that the signal of the ``phase-transition-probe'' {[}cf.~Eq.~(\ref{eq:classification-accuracy}){]}
is associated with a narrow parameter interval $[p_{0}^{*}-\Delta p_{0},p_{0}^{*}+\Delta p_{0}]$.
If there exists a strong phase coexistence in the certain parameter
region, as ubiquitous in the vicinities of the first-order phase transitions,
the probe indeed can also give the signal $\mathcal{P}(p_{0}^{*})\approx50\%$
in this parameter region despite possible first-order phase transitions
could exists. Therefore, the single peak in Fig.~\ref{fig:Identification-of-hexatic}(c)
might also indicate that there exists a relatively large phase coexistence
parameter region on its right. This motivates us to perform an auxiliary
machine learning, where we train the CNN using the data associated
with two parameter points that are well separated from each other.
More specifically, we use the data with $p_{0}=3.705$ and $p_{0}=3.800$,
and label all the samples associated with the former as ``phase A''
and the latter as ``phase B''. We then test the average confidence
values $(\mathcal{C}_{\textrm{A}},\mathcal{C}_{\textrm{B}})$ of the
trained CNN for $p_{0}\geq3.705$, where $\mathcal{C}_{\textrm{A}(\textrm{B})}\equiv\langle y_{\textrm{A}(\textrm{B})}\rangle$
and $\langle\cdot\rangle$ denotes the average over $M$ test samples
associated with each $p_{0}$. As we can see from the inset of Fig.~\ref{fig:Identification-of-hexatic}(c),
the average outputs $(\mathcal{C}_{\textrm{A}},\mathcal{C}_{\textrm{B}})$,
i.e., the average classification confidence values associated with
phase A and phase B, show big differences around $100\%$ in the vicinities
of both ends of the parameter interval $[3.705,3.800]$ for $p_{0}$.
This clearly indicates that there indeed exist two distinct phases
around $p_{0}=3.705$ and $p_{0}=3.800$, and thus the parameter region
around the intersection point in the inset of Fig.~\ref{fig:Identification-of-hexatic}(c)
is expected to correspond to a phase coexistence region. From the
expression of the system's elastic energy in Eq.~(\ref{eq:energy}),
we know that the systems with the small $p_{0}$ on the left side
of the peak are in the solid phase. While the intersection point in
the inset of Fig.~\ref{fig:Identification-of-hexatic}(c) appears
far right of the peak, so the phase coexistence around the intersection
point is not relevant to the solid phase, but should be a coexistence
between the hexatic and the liquid phases. Therefore, we can conclude
that the peak in Fig.~\ref{fig:Identification-of-hexatic}(c) corresponds
to the solid-hexatic phase transition point $p_{0}^{\textrm{SH}}=3.7025\pm0.0025$,
which is also consistent with the phase transition point $p_{0}^{\textrm{SH}}\approx3.69$
at $T=0.02$ identified via conventional approaches \citep{Ciamarra_PRM_2018},
and that the intersection point in the inset of Fig.~\ref{fig:Identification-of-hexatic}(c)
should correspond to an upper bound of the hexatic phase region with
$p_{0}^{\textrm{H,UB}}=3.7507$. Moreover, the existence of this relatively
large phase coexistence region naturally leads to the conclusion that
the hexatic-liquid phase transition in this system is discontinuous.

\emph{Critical scaling extracted via ``information-concealing''}.---So
far, we have identified the intermediate hexatic phase and clarified
the discontinuous nature of the hexatic-liquid phase transition. One
fundamental question still left is the critical scaling of the solid-hexatic
phase transition. To address this question, we develop a generic machine
learning approach dubbed ``information-concealing'' that is able
to extract the critical scaling of the correlation length in the vicinity
of a generic continuous phase transition.

This approach is based on the natural expectation that for a CNN employed
to perform a classification task in a translational invariant way,
its outputs, i.e., the classification confidence values, shall be
dominated by the features whose characteristic length scales are given
by the correlation length $\xi$ of the system. This thus indicates
that the content of the effective information of each sample of the
data, which the CNN can utilize to perform the classification task,
is determined by a characteristic ratio $\lambda\equiv l\slash\xi$
with $l$ being the length scale of the effective area of the sample
{[}cf.~Fig.~\ref{fig:Critical-scaling}(a){]}. Consequently, for
two different samples, as long as their characteristic ratios $\lambda$
are the same, their corresponding average confidence values of the
CNN are expected to be the same, too. This directly indicates that
for a generic continuous phase transition triggered by tuning a system
parameter $g$ across its critical value $g_{c}$, the average confidence
values as a function of the rescaled system parameter $(g-g_{c})l^{1\slash\nu}$
with different fixed $l$ should collapse to the same universal function
near the phase transition, since the correlation length $\xi$ generically
manifests the power law behavior $\xi\propto(g-g_{c})^{-\nu}$ in
the vicinity of the continuous phase transition with $\nu$ being
the critical exponent for the divergence of the correlation length.
Therefore, the critical exponent $\nu$ can be determined according
to the data collapse behavior of the average confidence values as
a function of the rescaled system parameter $(g-g_{c})l^{1\slash\nu}$
with different fixed $l$.

To implement the ``information-concealing'' approach, the information
of the samples in the original dataset are deliberately concealed
to generate a series of datasets with reduced effective areas for
their samples. Each of these samples with reduced effective areas
is obtained by randomly choosing an area of $l\times l$ pixels from
the original images and conceal its periphery with the white color
{[}cf.~Fig.~\ref{fig:Critical-scaling}(a){]}. The critical scaling
of the correlation length $\xi$ in the vicinity of the phase transition
under consideration is then straightforwardly obtained by investigating
the data collapse behavior of the average confidence values as a function
of the rescaled system parameter $(g-g_{c})l^{1\slash\nu}$ with different
fixed $l$ (see Supplemental Material \citep{Supplemental_Material}
for more technical details). A benchmark of this approach is presented
in Supplemental Material \citep{Supplemental_Material}, where a good
estimation of the critical exponent $\nu_{\textrm{I}}=1.04$ (exact
value $\nu_{\textrm{I}}=1$ \citep{Kogut_RMP_1988}) for the ferromagnetic-paramagnetic
transition of the 2D Ising model is obtained.

We directly apply this ``information-concealing'' approach to reveal
the critical scaling of the solid-hexatic phase transition in deformable
particles with the CNN employed to perform the classification tasks.
More specifically, we first perform ``concealing'' with a choice
of effective length scale $l$ to generate the corresponding dataset.
Then, this complete dataset is used to train the CNN to perform the
classification between the solid and the hexatic phases. After training,
the CNN is able to output the classification confidence values $(y_{\textrm{S}},y_{\textrm{H}})$
concerning the samples with corresponding target shape index $p_{0}$,
i.e., establish a map between the average confidence values $(\mathcal{C}_{\textrm{S}},\mathcal{C}_{\textrm{H}})$,
where $\mathcal{C}_{\textrm{S}(\textrm{H})}\equiv\langle y_{\textrm{S}(\textrm{H})}\rangle$,
and the target shape index $p_{0}$, from which the phase transition
point $p_{0}^{\textrm{SH}}$ is identified by the $p_{0}$ where $\mathcal{C}_{\textrm{S}}(p_{0})=\mathcal{C}_{\textrm{H}}(p_{0})$.
We repeat this process with a series of different choices of the effective
length scale $l=224,200,180,150,120,100$, and get the corresponding
$\mathcal{C}_{\textrm{S}}^{l}(p_{0})$ and $\mathcal{C}_{\textrm{H}}^{l}(p_{0})$.
Fig.~\ref{fig:Critical-scaling}(b) shows the dependence of the average
confidence values on $p_{0}$ with different fixed $l$ at the low
temperature $T=0.005$, where the behaviors differ with respect to
the target shape index $p_{0}$. However, $\mathcal{C}_{\textrm{S}(\textrm{H})}^{l}$
indeed shows the same dependence on its rescaled target shape index
$(p_{0}-p_{0}^{\textrm{SH}})l^{1\slash\nu}$ with $\nu=0.64$, as
shown in Fig.~\ref{fig:Critical-scaling}(c), indicating that the
critical exponent $\nu$ for the divergence of the correlation length
of the solid-hexatic phase transition assumes the value $\nu=0.64$.
We further perform the similar investigation at the relatively high
temperature $T=0.02$, the key results of which are shown in Figs.~\ref{fig:Critical-scaling}(d),~(e),
where we notice that the good data collapse for the average confidence
values is observed for $\nu=0.66$. This is consistent with the value
of the critical exponent obtained at $T=0.005$, indicating that the
critical scaling of the correlation length $\xi$ in the vicinity
of the solid-hexatic phase transition is given by the power law $\xi\propto(p_{0}-p_{0}^{\textrm{SH}})^{-\nu}$
with $\nu\approx0.65$.

Finally, we remark that both ``scanning-probe'' and ``information-concealing''
machine learning involve no special design of the CNN structure. A
widely-used standard 18-layer residual neural network \citep{Kaiming_He_Proceedings_IEEE_CVPR_2016}
is employed to perform the analysis as shown in Fig.~\ref{fig:Identification-of-hexatic}
and Fig.~\ref{fig:Critical-scaling} in the main text. In Supplemental
Material \citep{Supplemental_Material}, we present the same analysis
employing another type of deep CNN called ``GoogLeNet'' \citep{Szegedy_IEEE_CVPR_2015},
which gives the same results as the ones obtained by the residual
neural network. This clearly shows the CNN structure independence
of the two approaches developed in this work, which is indispensable
for the reliability of the physical results they predict. Moreover,
this also indicates the wide generic applicability of these approaches
in investigating other relevant complex systems.

\emph{Conclusion and outlook}.---By directly analyzing system's spatial
configurations via ``scanning-probe'' and ``information-concealing''
machine learning developed in this work, we provide machine learning
evidence for the existence of the intermediate hexatic phase in 2D
interacting deformable polymeric particles, and obtained the critical
scaling behavior of the solid-hexatic phase transition, where the
divergence of the correlation length is determined by a power law
with the critical exponent $\nu\approx0.65$. Since these two machine
learning approaches involve no special design for the CNN employed,
they can be readily employed as a new generic toolbox to investigate
the existence of the intermediate phase and the possible critical
scaling behavior in various complex systems. For instance, we believe
that our work will stimulate further efforts in revealing the critical
scaling behavior of the solid-hexatic phase transition in nonequilibrium
complex systems, e.g., in self-propelled biological tissues, via these
new approaches. Moreover, noticing that the deep CNNs employed here
are in general quite powerful in pattern recognition, we also expect
that our machine learning approaches can be applied to investigate
the physics associated with 2D microphase separations \citep{Liebchen_PRL_2017,Caporusso_PRL_2020,Ran_Ni_Sci_Adv_2019,Ran_Ni_JChemPhys_2022}
in various complex systems.
\begin{acknowledgments}
We thank Jiajian Li for useful discussions. This work was supported
by NSFC (Grant No.~11874017 and No.~12075090), GDSTC (Grant No.~2018A030313853
and No.~2017A030313029), GDUPS (2016), Major Basic Research Project
of Guangdong Province (Grant No.~2017KZDXM024), Science and Technology
Program of Guangzhou (Grant No.~2019050001), and START grant of South
China Normal University.
\end{acknowledgments}

\end{document}

\renewcommand{\theequation}{S-\arabic{equation}} \renewcommand{\thefigure}{S\arabic{figure}} \renewcommand{\thetable}{S\arabic{table}}
\title{Supplemental Material for ``Scanning-probe and information-concealing
machine learning intermediate hexatic phase and critical scaling of
solid-hexatic phase transition in deformable particles''}
\author{Wei-chen Guo}
\affiliation{Guangdong Provincial Key Laboratory of Quantum Engineering and Quantum
Materials, School of Physics and Telecommunication Engineering, South
China Normal University, Guangzhou 510006, China}
\affiliation{Guangdong-Hong Kong Joint Laboratory of Quantum Matter, South China
Normal University, Guangzhou 510006, China}
\author{Bao-quan Ai}
\email{aibq@scnu.edu.cn}

\affiliation{Guangdong Provincial Key Laboratory of Quantum Engineering and Quantum
Materials, School of Physics and Telecommunication Engineering, South
China Normal University, Guangzhou 510006, China}
\affiliation{Guangdong-Hong Kong Joint Laboratory of Quantum Matter, South China
Normal University, Guangzhou 510006, China}
\author{Liang He}
\email{liang.he@scnu.edu.cn}

\affiliation{Guangdong Provincial Key Laboratory of Quantum Engineering and Quantum
Materials, School of Physics and Telecommunication Engineering, South
China Normal University, Guangzhou 510006, China}
\affiliation{Guangdong-Hong Kong Joint Laboratory of Quantum Matter, South China
Normal University, Guangzhou 510006, China}
\begin{abstract}
This Supplemental Material provides (i) technical information about
the data generation and the network training, (ii) a benchmark of
the ``information-concealing'' approach, (iii) results that show
the CNN structure independence of both approaches developed in this
work, and (iv) investigations on influences of the finite size effect
and the number of samples in the dataset.
\end{abstract}
\maketitle
\begin{figure}
\noindent \begin{centering}
\includegraphics[width=3.2in]{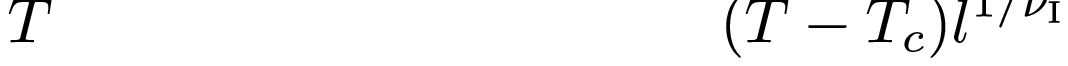}
\par\end{centering}
\caption{\label{fig:Ising-benchmark}(a) Average confidence values of the CNN
to classify the ferromagnetic phase from the paramagnetic phase using
datasets with different fixed $l$ of the 2D Ising model without external
field. (b) The good data collapse is achieved for the average confidence
values in (a) as a function of $(T-T_{c})l^{1\slash\nu_{\textrm{I}}}$
with the critical exponent $\nu_{\textrm{I}}=1.04$. This is a good
estimation of the critical exponent for the divergence of the correlation
length of the 2D Ising model (exact value $\nu_{\textrm{I}}=1$).
See text for more details.}
\end{figure}

\begin{figure}
\noindent \begin{centering}
\includegraphics[width=3.2in]{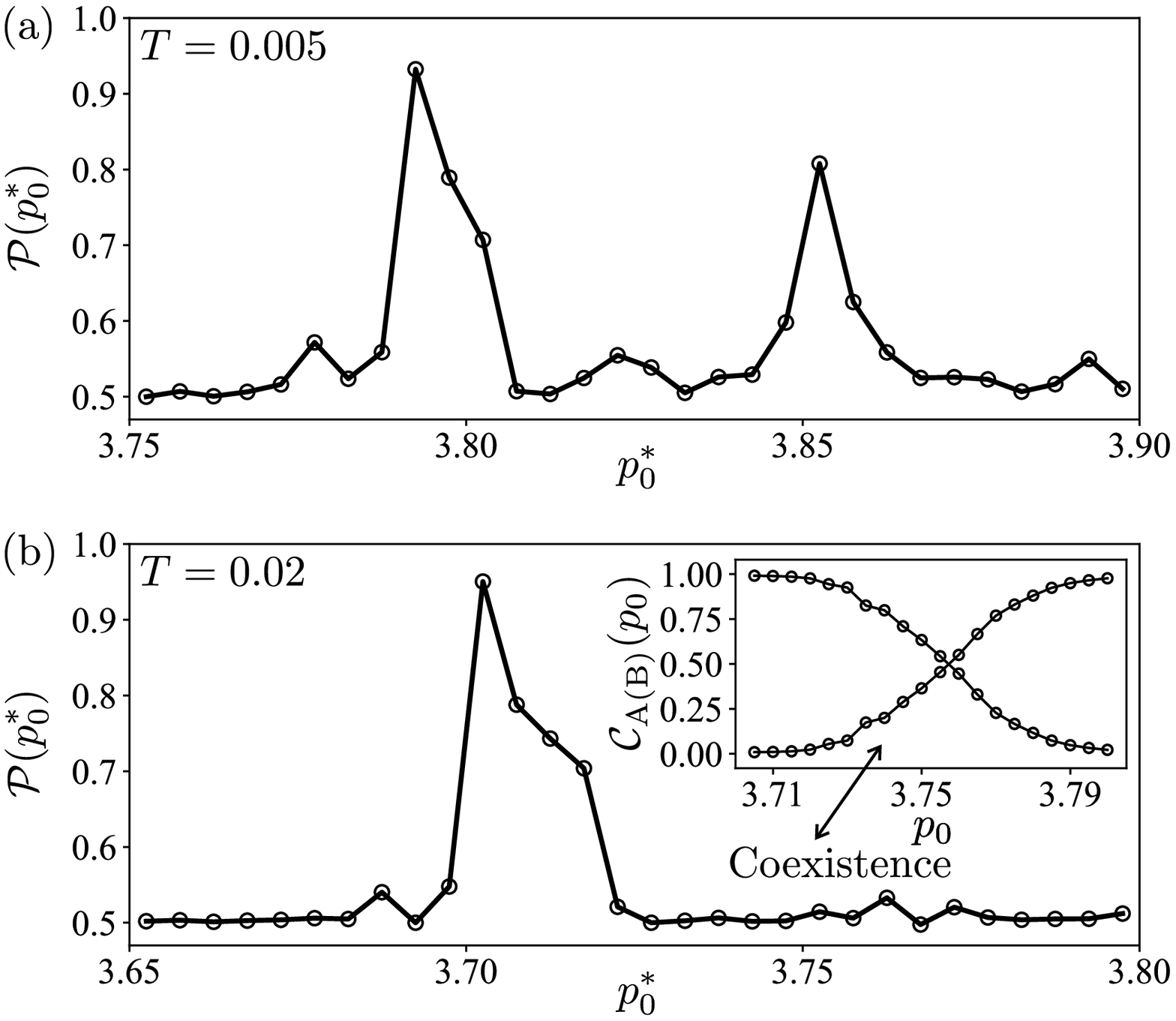}
\par\end{centering}
\caption{\label{fig:Identification-of-hexatic-ggn}Analogous data to Figs.~1(b),~(c),
but employing another type of deep CNN called ``GoogLeNet''. These
results match very well with the ones obtained by the residual neural
network. See text for more details.}
\end{figure}

\section*{Data generation and network training}

In this work, for each $p_{0}=3.750,3.755,\ldots,3.900$ at the low
temperature $T=0.005$, and $p_{0}=3.650,3.655,\ldots,3.800$ at the
relatively high temperature $T=0.02$, we perform Brownian dynamics
simulations for $10^{6}$ time steps with $dt=0.01$ to equilibrate
the system, and confirm it from the convergence of the system's elastic
energy. Afterwards, the dynamics of the system is simulated for additional
$10^{5}$ time steps, during which the spatial configuration of the
system is sampled every 50 time steps. These configurations are transformed
to images of $589\times442$ pixels, and divided into three categories
according to the usages of machine learning, namely, training set
($8\times10^{2}$ samples for each $p_{0}$), validation set ($2\times10^{2}$
samples for each $p_{0}$), and test set ($10^{3}$ samples for each
$p_{0}$). The input layer of the deep CNN employed has $3\times224\times224$
neurons (3 for RGB, and $224\times224$ is a standard input size),
and the output layer has 2 neurons, i.e., $y_{\textrm{A}}$ and $y_{\textrm{B}}$.
The CNN is trained in both ``scanning-probe'' and ``information-concealing''
machine learning as a binary classifier that can extract the distinguishable
features from the system's spatial configurations and predict the
class of the input samples. This prediction is compared to the given
label of the sample and the error can be quantified by the cross-entropy
cost function \citep{Nielsen_Book_2015,Goodfellow_Book_2016}. The
training of a CNN is realized by minimizing the cost function traversing
the training set for several epochs. More specifically, the CNN is
trained with the Adam optimization algorithm \citep{Kingma_arXiv_2014}
starting from the pre-trained model using default settings in PyTorch.
The auxiliary machine learning to resolve the hexatic-liquid phase
coexistence is obtained by training and validating for 50 epoch traversing
the samples and all the other results are obtained by training and
validating for 5 epoch, repeating independently for 20 times to take
the average. For more thorough discussions on the machine learning
techniques involved in the training process, we refer the reader to
Refs.~ \citep{Nielsen_Book_2015,Goodfellow_Book_2016}.

\begin{figure}
\noindent \begin{centering}
\includegraphics[width=3.2in]{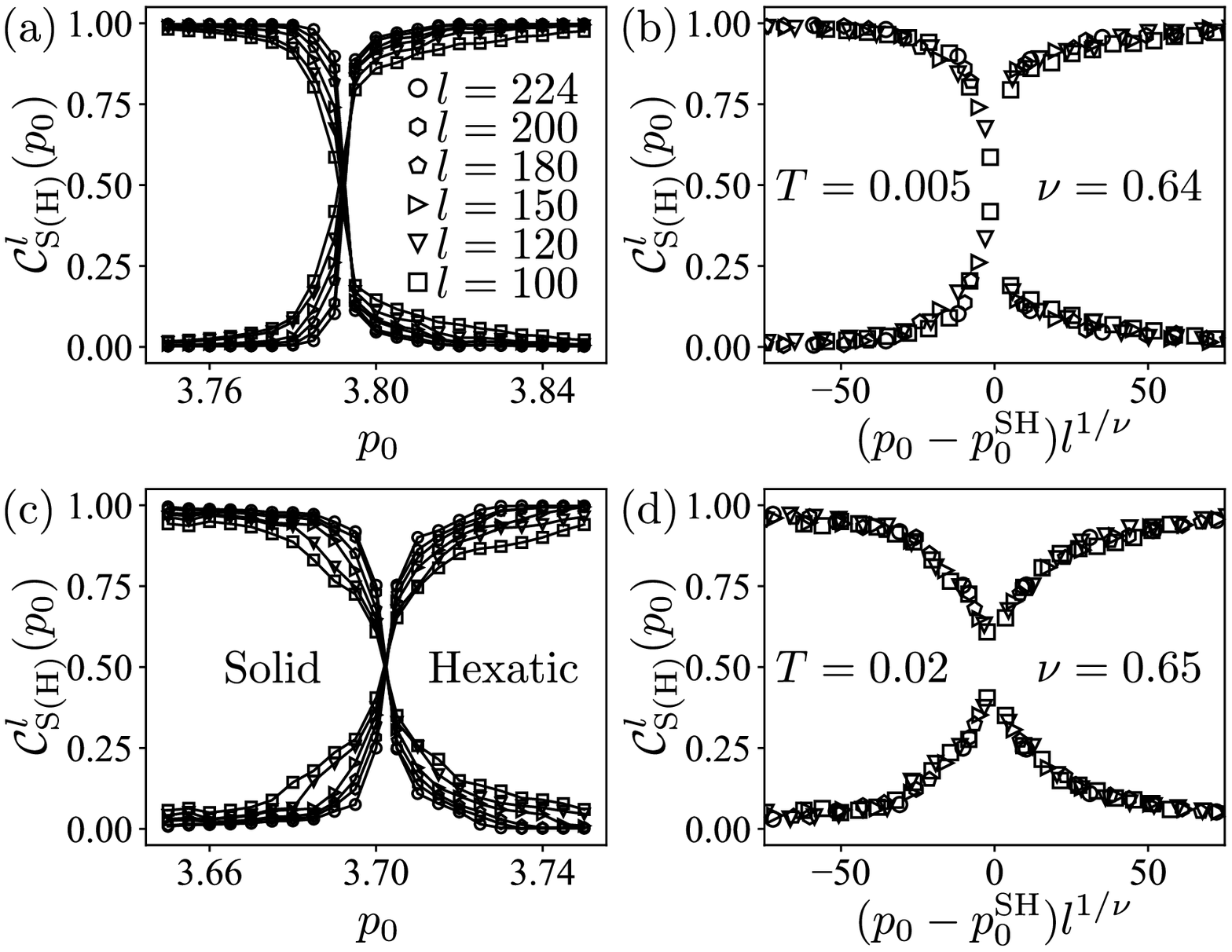}
\par\end{centering}
\caption{\label{fig:Critical-scaling-ggn}Analogous data to Figs.~2(b-e),
but employing another type of deep CNN called ``GoogLeNet''. These
results also match very well with the ones obtained by the residual
neural network. See text for more details.}
\end{figure}

\section*{Benchmark of ``information-concealing''}

The ``information-concealing'' approach developed in this work is
inspired by Refs.~\citep{Melko_Nat_Phys_2017,van_Nieuwenburg_arXiv_2021},
making use of the CNN to analyze the system's spatial configurations
from numerical simulations with particle number $N$ fixed. Here,
we present a benchmark of the ``information-concealing'' approach.
We directly apply this approach to reveal the critical scaling of
the correlation length in the vicinity of the ferromagnetic-paramagnetic
transition of the 2D Ising model with the standard 18-layer residual
neural network \citep{Kaiming_He_Proceedings_IEEE_CVPR_2016} employed
to perform the classification tasks. The samples are generated from
Monte Carlo simulations at different temperatures $T$ in a square
lattice of linear size $L=224$ with periodic boundary condition imposed
and without external field. Similar to what we have done in the main
text, we first perform ``concealing'' with a choice of effective
length scale $l$ to generate the corresponding dataset. Then, this
complete dataset is used to train the CNN to perform the classification
between the ferromagnetic and the paramagnetic phases. After training,
the CNN is able to output the classification confidence values $(y_{\textrm{F}},y_{\textrm{P}})$
concerning the samples with corresponding temperature $T$, i.e.,
establish a map between the average confidence values $(\mathcal{C}_{\textrm{F}},\mathcal{C}_{\textrm{P}})$,
where $\mathcal{C}_{\textrm{F}(\textrm{P})}\equiv\langle y_{\textrm{F}(\textrm{P})}\rangle$,
and the temperature $T$, from which the ferromagnetic-paramagnetic
transition point, i.e., the so-called Curie temperature $T_{c}$,
is identified by the $T$ where $\mathcal{C}_{\textrm{F}}(T)=\mathcal{C}_{\textrm{P}}(T)$.
We repeat this process with a series of different choices of the effective
length scale $l=224,200,180,150,120,100$, and get the corresponding
$\mathcal{C}_{\textrm{F}}^{l}(T)$ and $\mathcal{C}_{\textrm{P}}^{l}(T)$.
The average confidence values with different fixed $l$ show the same
dependence on its rescaled temperature $(T-T_{c})l^{1\slash\nu_{\textrm{I}}}$
with $\nu_{\textrm{I}}=1.04$ in Fig.~\ref{fig:Ising-benchmark},
indicating that the critical exponent $\nu_{\textrm{I}}$ for the
divergence of the correlation length of the ferromagnetic-paramagnetic
transition assumes the value $\nu_{\textrm{I}}=1.04$, matching well
with the exact value $\nu_{\textrm{I}}=1$ \citep{Kogut_RMP_1988}.

\begin{figure}
\noindent \begin{centering}
\includegraphics[width=3.2in]{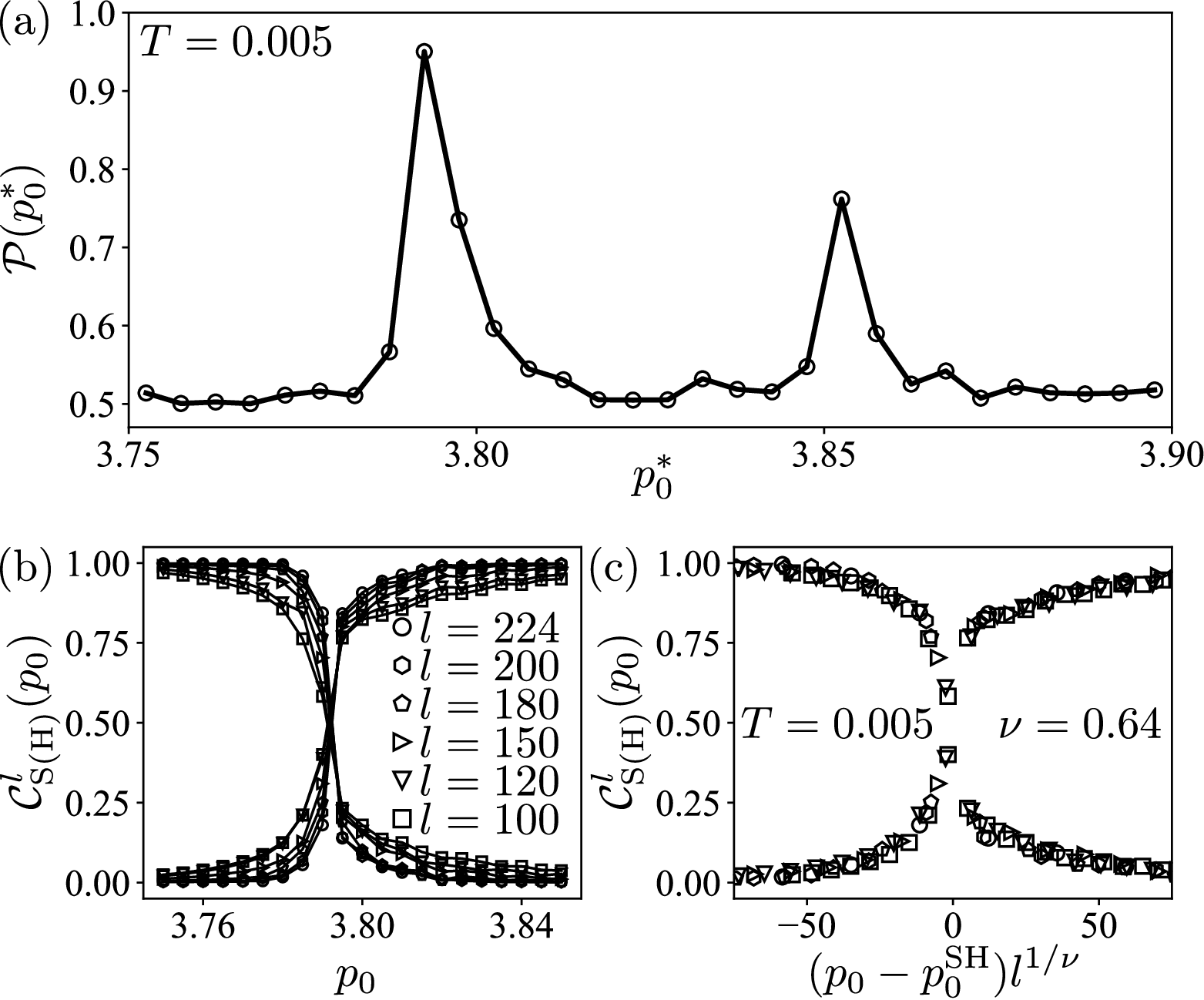}
\par\end{centering}
\caption{\label{fig:Small-dataset}Analogous data to Figs.~1(b),~2(b), and
2(c), but using only half of the samples, i.e., only $10^{3}$ for
each $p_{0}$ at the low temperature$T=0.005$. These results suggest
that the number of samples used in the main text, i.e., $2\times10^{3}$
for each $(T,p_{0})$, is large enough. See text for more details.}
\end{figure}

\begin{figure}
\noindent \begin{centering}
\includegraphics[width=3.2in]{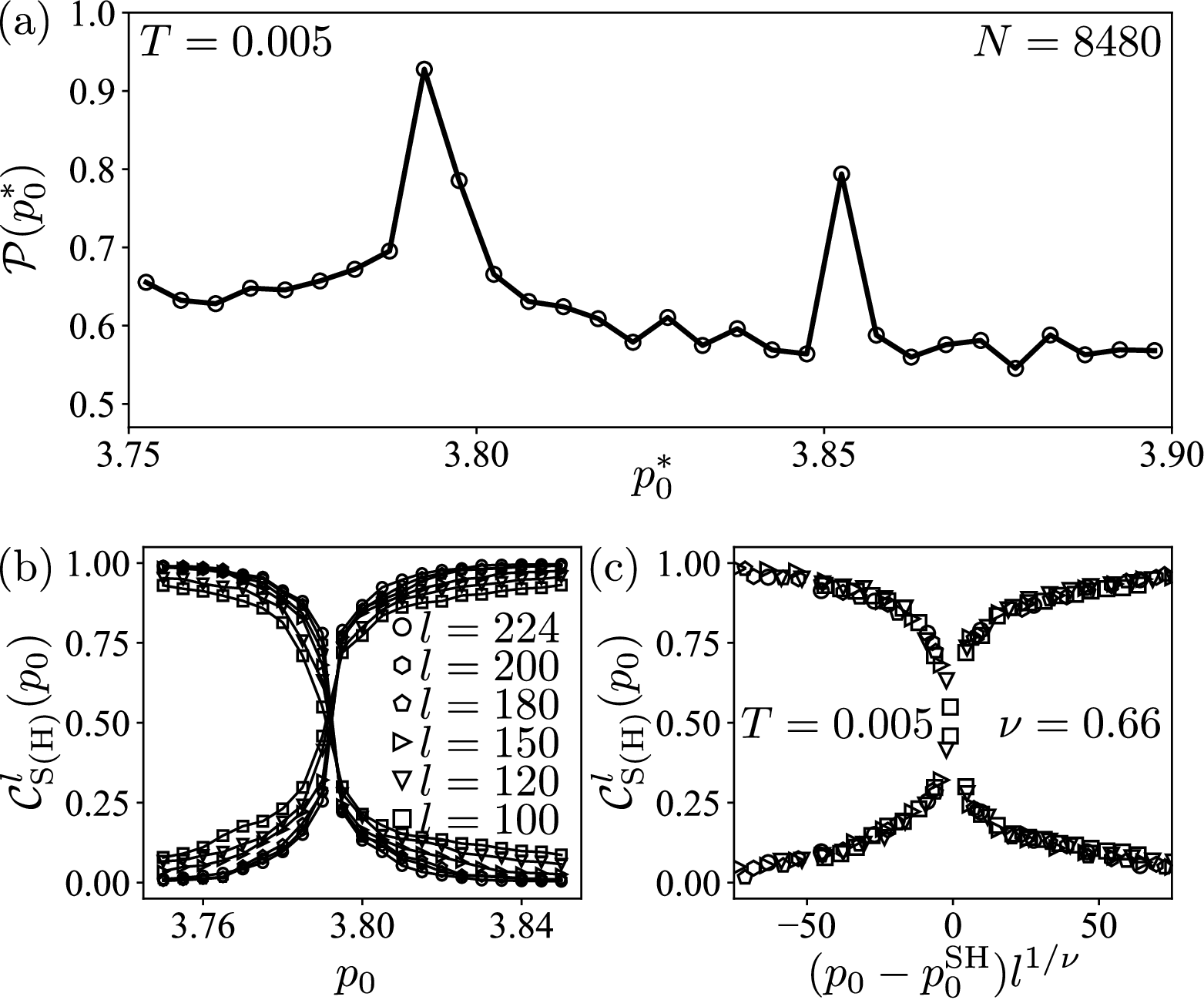}
\par\end{centering}
\caption{\label{fig:Finite-size-effect}Analogous data to Figs.~1(b),~2(b),
and 2(c), but considering a larger system size $N=8480$. These results
show that the finite size effect does not impose strong influences
on the major results presented in the main text. See text for more
details.}
\end{figure}

\section*{Independence on the CNN structure}

In the main text, we employ a standard 18-layer residual neural network
\citep{Kaiming_He_Proceedings_IEEE_CVPR_2016} to perform the analysis
as shown in Fig.~1 and Fig.~2. Here, we present the same analysis
employing another type of deep CNN called ``GoogLeNet'' \citep{Szegedy_IEEE_CVPR_2015}.
As we can see from Fig.~\ref{fig:Identification-of-hexatic-ggn},
the ``phase-transition-probe'' realized by ``GoogLeNet'' detects
the same phase transition points as the ones obtained by the residual
neural network, and the upper bound for the hexatic phase region is
estimated by the intersection point with $p_{0}^{\textrm{H,UB}}=3.7573$
in the inset of Fig.~\ref{fig:Identification-of-hexatic-ggn}(b).
Fig.~\ref{fig:Critical-scaling-ggn} shows the results of revealing
the critical scaling of the solid-hexatic phase transition in deformable
particles with ``GoogLeNet'' employed to perform the classification
tasks, which also matches very well with ``information-concealing''
machine learning results obtained by the residual neural network in
Fig.~2. We remark that these two deep CNNs have quite different structures,
so the good consistency of the results clearly shows the CNN structure
independence of both approaches developed in this work.

\section*{Influences of the finite size effect and the number of samples in
the dataset}

At a given temperature $T$, we have $6.2\times10^{4}$ samples in
total. To investigate the influences of the number of samples in the
dataset, here we use half of the samples, i.e., only $10^{3}$ for
each $p_{0}$, to perform the same analysis at the low temperature
$T=0.005$. As we can see in Fig.~\ref{fig:Small-dataset}(a), the
``scanning-probe'' machine learning identifies the solid-hexatic
phase transition point $p_{0}^{\textrm{SH}}=3.7925\pm0.0025$ and
the hexatic-liquid phase transition point $p_{0}^{\textrm{HL}}=3.8525\pm0.0025$,
which are exactly the same as the ones obtained in Fig.~1(b) and
Fig.~\ref{fig:Identification-of-hexatic-ggn}(a). And the ``information-concealing''
machine learning extracts the critical exponent $\nu=0.64$ in Fig.~\ref{fig:Small-dataset}(c)
for the divergence of the correlation length, which is also the same
as the one obtained in Fig.~2(c) and Fig.~\ref{fig:Critical-scaling-ggn}(b).
These results show that reducing the number of samples in the dataset
by half from $6.2\times10^{4}$ to $3.1\times10^{4}$ does not impose
noticeable influences on the major results, suggesting that the number
of samples used in the main text, i.e., $2\times10^{3}$ for each
$(T,p_{0})$, is large enough. 

Moreover, we also investigate the influences of the finite size effect,
since it can be generally important in the melting transition of 2D
systems \citep{Ran_Ni_Commun_Phys_2019}. Here, we consider a larger
system size $N=8480$, and find that the predicted phase transition
points $p_{0}^{\textrm{SH}}$ and $p_{0}^{\textrm{HL}}$ are unchanged
compared with the ones obtained in the main text {[}cf.~Fig.~\ref{fig:Finite-size-effect}(a)
and Fig.~1(b){]}. The critical exponent extracted via ``information-concealing''
machine learning is $\nu=0.66$, which also agrees well with the results
obtained in the main text {[}cf.~Fig.~\ref{fig:Finite-size-effect}(c)
and Fig.~2(c){]}. These results show that the finite size effect
does not impose strong influences on the major results presented in
the main text.